\documentclass[journal,a4paper]{IEEEtran}
\usepackage{graphicx,dsfont,amssymb,amsmath,amsthm,cite,calc,tabulary,setspace}
\usepackage[caption=false,font=footnotesize]{subfig}
\usepackage{xcolor}
 \usepackage{algorithm,algpseudocode}
\usepackage{algpseudocode}
\usepackage{mathrsfs}

\usepackage{bbm}

\usepackage{color,soul}

\begin{document}
\title{Fast Decentralized Linear Functions via Successive Graph Shift Operators}

\author{Siavash Mollaebrahim, Daniel~Romero,~\IEEEmembership{Member,~IEEE,}
       and~Baltasar~Beferull-Lozano,~\IEEEmembership{Senior Member,~IEEE}
\thanks{The authors are with the Intelligent Signal Processing and Wireless Networks (WISENET) center, Department of Information and Communication Technology, University of Agder, Norway. e-mail:\{ siavash.mollaebrahim, daniel.romero, baltasar.beferull\}@uia.no}
\thanks{This work was supported by the PETROMAKS Smart-Rig grant 244205/E30, the SFI Offshore Mechatronics grant 237896/O30, the IKTPLUSS INDURB grant 270730/O70 from the Research Council of Norway.}
}
\maketitle
\begin{abstract}
We study decentralized designing of the graph shift operators to implement linear transformations between graph signals. Since this operator captures the local structure of the graph, the proposed method of this paper gives rise to decentralized linear network operators. Unfortunately, existing decentralized approaches either consider some special instances of linear transformations or confine themselves to some known graph shift operators reduced family of the designing linear transformations task. To
remedy these limitations, this paper develops a framework for computing a wide class of linear transformations in a decentralized fashion by relying on the notion of graph shift operator. To this end, a set of successive graph shift operators is implemented to compute linear transformations in a small number of iterations (as fast as possible). 
\end{abstract}

\section{introduction}
The field of graph signal processing (GSP) enables us to perform many different tasks such as denoising, compression or reconstruction on signals defined on graphs. Redesigning the traditional signal processing tools (used to process signals defined on regular domains) to be able to process signals on more complex graph domain is critical. Graph filters (GFs) are a tool to accomplish those processes distributively. GFs can be expressed by matrix polynomials of a local operator called the graph-shift~\cite{sandryhaila2014discrete}. 

One of the important applications of GFs is subspace projection. Indeed, a number of learning tasks in wireless sensor networks (WSN) such as estimation and denoising, can be done by projecting the observed signal onto a known subspace~\cite{barbarossa2009projection}. One way to carry out task is using a centralized method in which all sensors gather their local information, and then send it to a central processor where the subspace projection is calculated. However, this approach needs a large number of transmissions to relay the sensor data to the central processor, which incurs high energy consumption and increases the cost of sensor hardware (since they are usually battery-powered) making the approach unfeasible. 

In the centralized approach, sensor nodes near the central processor should relay its own packet and the other further nodes packets, thus using more energy which shortens their lifetime. The consequence of this issue is that the central processor gets disconnected from the rest of the network. Consequently, decentralized approaches with no central processor where subspace projection is performed by only exchanging local information between nodes is vital.

To end this task, a decentralized subspace projection method has been proposed in~\cite{barbarossa2009projection} where each node at every iteration linearly combines its iterate with the ones of its neighbors. Then, in order to obtain the coefficients of those linear combinations, a criterion that quantifies asymptotic convergence is optimized. However, the convergence of this method is asymptotic, and needs a sufficiently large number of iterations (local information exchanges). This restriction has been addressed in \cite{sandryhaila2014finitetime}, \cite{segarra2017operators}, \cite{safavi2015nulling} via graph filters for a special case of subspace projection (average consensus). 

The advantage of these methods is that they are capable of converge in a finite number of iterations. In \cite{segarra2017operators}, the authors consider the case beyond subspace projection,  where they design GFs to compute pre-specified linear transformations over the graph signals. Nevertheless, \cite{segarra2017operators} restricts itself to design GFs for rank-1 projections or to projections that share eigenvectors with given shift matrices (the Laplacian or adjacency matrix). 

To address those limitations, in~\cite{romero2018projection}, the graph shift operator is designed by optimizing a criterion that yields convergence to the subspace projection in a nearly minimal number of iterations. However, the proposed method is not appropriate for large networks because directly optimizing that criterion, involves a complexity of $\mathcal{O}(N^6)$, where $N$ is the number of sensor nodes. Finally, this issue has been alleviated  in ~\cite{mollaebrahim2018projection} by reformulating the aforementioned criterion and using an optimization algorithm with complexity $\mathcal{O}(N-r)^{3}$, where $r$ is the subspace projection dimension.  Therefore, the method in ~\cite{mollaebrahim2018projection} could consider a broader range of scenarios (larger networks) in comparison with~\cite{romero2018projection}, and provide subspace projection in a nearly minimal number of iterations. 

 In this paper, to obtain any pre-specified linear transformation, an optimization problem is proposed which consists of finding a set of successive graph shift operators. The proposed method aims to get the linear transformations as fast as possible (in a finite number of iterations) at the expense of using a sequence of different graph shift operators.  In addition, it can consider larger scenarios compared to the other works in the context of GSP such as ~\cite{romero2018projection}, ~\cite{mollaebrahim2018projection}, \cite{segarra2017operators} and ~\cite{barbarossa2009projection}. In fact, the approach is applicable for designing any arbitrary pre-specific linear transformations, and various topologies (symmetric and asymmetric graph networks).       

The remainder of the paper is structured as follows. Sec. 2 introduces notation and reviews some existing results on decentralized subspace projection with graph filters. Sec. 3 presents the proposed algorithm. Finally, Sec. 4 validates its performance through numerical experiments and Sec. 5 concludes the paper.

\text{Notation}: $\mathbf{v}_{n}$ and $\mathbf{A}_{n,n'}$ denote the $n$-th entry of $\mathbf{v}$ and the entries of $\mathbf{A}$, respectively. Also, $\mathbf{1}$ is the ones vector and $\mathbf{e}_{n}$ is the basis vector whose all entries are zero except the $n$-th one which equals one. 
 
\section{Preliminaries and Problem formulation}
 Let a directed connected network having $N$ sensor nodes be represented by $\mathcal G(\mathcal V,\mathcal E)$, where $\mathcal V=\{v_1,\ldots,v_N\}$ is the set of vertices and $\mathcal E \subset \mathcal V \times \mathcal V $ the set of edges. The $n'$-th vertex $v_{n'}$ is connected to the $n$-th vertex $v_{n}$ if there is a directed edge from $v_{n'}$ to $v_{n}$ $(v_{n'},v_{n})\in\mathcal E$ (this does not mean that $v_{n}$ is connected to $v_{n'}$ unless $(v_{n},v_{n'})\in\mathcal E$).  
  The in-neighborhood of the $n$-th node is defined as the set of nodes connected to it, which can be denoted as $\mathcal{N}_{n}=\{v_{n'}\mid(v_{n'},v_n)\in\mathcal{E}\}$.

The input signal (for instance in WSN, it is formed by measurements collected by sensors) is $\mathbf{z}\in\mathbb{R}^{N}$. The goal of the paper is to find $\mathbf{H}\mathbf{z}$ ( in a decentralized fashion ) given $\mathbf{H}$, $\mathbf{z}$ and $\mathcal G$, where $\mathbf{H}$ is a linear transformation matrix.  

One idea to reach that goal is using the graph shift operator which is an operator $\mathbf{z}\mapsto\mathbf{S}\mathbf{z}$ where $\mathbf{S}\in\mathbb{R}^{N\times{N}}$ satisfies $\mathbf{S}_{nn'}=0$ if $(v_{n'},v_{n})\not\in\mathcal E$. The graph shift operator is a decentralized operator because $(\mathbf{S}\mathbf{z})_{n}=\sum_{n':v_{n'}\in\mathcal{N}_{n}} (\mathbf{S})_{nn'}\mathbf{z}_{n'}$ which means that the $n$-th node computes the summation only with information of its own neighbours. 

The graph filters are linear graph-signal operators $\mathbf{H}:\mathcal{R}^{N}\mapsto\mathcal{R}^{N}$ of the form ${\mathbf H}:=\sum_{l=0}^{L-1}c_{l}{\mathbf S}^{l}$ ~\cite{Shuman2013GSP}.  The filter coefficients $\mathbf{c}^{\top}=[c_{1},c_{2},\cdots,c_{L-1}]$ have been used to define the graph filter $(\mathbf{H})$ as polynomials of $\mathbf{S}$ such that. In fact, the coefficients give a linearity to weight the contribution of each term $\mathbf{S}^{l}$ in the graph filter. 

From this, the graph shift operators form the basis for GFs. Indeed, graph filters can be implemented with $L-1$ exchanges of information between nodes. Let $\mathbf{z}\in\mathbb{R}^{N}$ the input graph signal, the $n$-th sensor exchanges its value $\mathbf{z}[n]$ to its neighbors. After exchanging the values $\mathbf{z}[n] \quad\forall {n=1,\cdots,N}$, then all nodes update their information via $\mathbf{z}^{(1)}=\mathbf{S}\mathbf{z}$. For a GF of order $L$, this procedure repeats for $L-1$ iterations. 

 Designing linear transformation functions includes a broad range of signal processing tasks such as the average consensus problem and the subspace projection task.
 In the average consensus problem, nodes seek to converge their values to the average of their initial values in a distributed fashion. Thus, in this case $\mathbf{H}=\frac{1}{N}(\mathbf{1}^{\top}\mathbf{1})$. In ~\cite{xiao2004average}, the authors have shown that the problem of finding edge weights for maximizing the asymptotic consensus speed can be expressed by a convex optimization problem.

  Another instance of designing linear transformation functions is subspace projection. In this task, $\mathbf{z}=\mathbf{x}+\mathbf{v}$, where $\mathbf{x}\in\mathbb{R}^{N}$ and $\mathbf{v}\in\mathbb{R}^{N}$ are the useful signal and the observation noise, respectively. The useful signal typically belongs to a subspace with dimension $r$ much smaller than $N$ i.e. $r<<N$.  Therefore, $\mathbf{x}$ can be expressed by $\mathbf{x}=\mathbf{U}_{\parallel}\boldsymbol{\alpha}$, where $\mathbf{U}_{\parallel}\in\mathbb{R}^{N\times{r}}$ is a matrix whose columns, assumed orthonormal w.l.o.g., span that subspace. The orthogonal projection of $\mathbf{z}$ onto the subspace spanned by the columns of $\mathbf{U}_{\parallel}$, which equals the least-squares estimate of $\mathbf{x}$, is given by $\hat{\mathbf{x}}=\mathbf{U}_{\parallel}\mathbf{U}_{\parallel}^{\top}\mathbf{x}=\mathbf{P}\mathbf{x}$, where $\mathbf{P}\overset{\Delta}{=}\mathbf{U}_{\parallel}\mathbf{U}_{\parallel}^{\top}$ is the projection matrix. Consequently, in this case $\mathbf{H}=\mathbf{P}$. 

 In the context of GSP,  recently a number of approaches have been proposed to design decentralized linear transformations such as \cite{segarra2017operators},~\cite{romero2018projection}, but still there are some challenges that should be addressed. For instance, ~\cite{romero2018projection} is just applicable to the symmetric topologies. Also, it can only be applied when $\mathbf{H}$ is the projection matrix. In \cite{segarra2017operators}, the graph shift operators have been confined to the Laplacian or adjacency matrix. Therefore, for the linear transformations that do not share their eigenvectors with the Laplacian or adjacency matrix, \cite{segarra2017operators} needs knowledge of the graph shift operator which is seldom known.      

 \section{Proposed method}
To address these issues, this section proposes a fast decentralized method via graph shift operators to compute $\mathbf{H}\mathbf{z}$. In this approach, a sequence of graph shift operators is applied to compute $\mathbf{S}_{L}\cdots\mathbf{S}_{2}\mathbf{S}_{1}\mathbf{z}$. From section II it follows that the proposed approach can be computed in a decentralized manner. After the first round of information exchange among nodes, the nodes compute $\mathbf{z}^{(1)}=\mathbf{S}_{1}\mathbf{z}$. Then, at the second round, $\mathbf{z}^{(2)}=\mathbf{S}_{2}\mathbf{z}^{(1)}=\mathbf{S}_{2}\mathbf{S}_{1}\mathbf{z}$. That procedure repeats for $L$ iterations. If $\{\mathbf{S}_{l}\}_{l=1}^{L}$ are properly designed, at the end of the algorithm, we have $\mathbf{H}\mathbf{z}\approx\Pi_{l=1}^{L}\mathbf{S}_{l}\mathbf{z}$ with a reasonable amount of error. 

This approach lets us design the graph shift operator to compute linear transformations for various possibly directed topologies. Also, similar to \cite{romero2018projection}, \cite{segarra2017operators}, it needs memory to store the graph shift operators. Suppose that the number of in-neighbours nodes of the $n$-th node is $E_{n}$. 

In \cite{romero2018projection}, \cite{segarra2017operators}, the $n$-th node should store $E_{n}+L'$ scalers (since in those methods there exists just one graph shift operator), where $L'$ is the order of the filter (for storing the filter coefficients).  However, in the proposed method, we have $L$ graph shift operators; thus, the $n$-th node needs to store $LE$ scalers.

One idea to obtain $\{\mathbf{S}_{l}\}_{l=1}^{L}$  is minimizing the error of the last iteration ${{\left\| \mathbf{H}\mathbf{z}-\Pi_{i=1}^{l}\mathbf{S}_{i}\mathbf{z} \right\|}_{F}^{2}}$ . However, as we stated before, the main goal of the proposed method is to compute $\mathbf{H}$ in the smallest number of iterations. This goal can be expressed by
 \begin{subequations}\label{goal}
  \begin{align}
  \underset{\mathbf{S}_{1},\mathbf{S}_{2},\cdots,\mathbf{S}_{L},l}{\text{min.}} &
       l\\
\text{s. t.}   \;\;\;\;    	&  (\mathbf S_{i})_{n,n'}=0 \hspace{2mm} \nonumber\\&\text{if} \hspace{2mm} (v_n,v_{n'})\not \in \mathscr E ,  n,n' =1,....,N, i=1,...,L\\
& \mathbf{H}=\Pi_{i=1}^{l}\mathbf{S}_{i} \\
& l\in\{1,\cdots,L\}
\end{align}
\end{subequations}
\eqref{goal} can be rewritten by using the indicator function as follows 
 \begin{subequations}\label{goal2}
  \begin{align}
  \underset{\mathbf{S}_{1},\mathbf{S}_{2},\cdots,\mathbf{S}_{L}}{\text{min.}} &
       \sum_{l=1}^{L} \mathbbm{1}(\mathbf{H}\neq\Pi_{i=1}^{l}\mathbf{S}_{i})\\
\text{s. t.}   \;\;\;\;    	&  (\mathbf S_{i})_{n,n'}=0 \hspace{2mm} \nonumber\\&\text{if} \hspace{2mm} (v_n,v_{n'})\not \in \mathscr E ,  n,n' =1,....,N, i=1,...,L
\end{align}
\end{subequations}
where $\mathbbm{1}$ denotes the indicator function and  $L$ is the maximum number of iterations allowed. 

Due to the fact that $\mathbf{S}$ is a local operator, the topology constraint is added to the problem. Moreover, if $\mathbf{S}$ must be symmetric, then $\mathbf{S}^{\top}=\mathbf{S}$ should be added as a constraint to the optimization problem. Note that if $\exists\{\mathbf{S}_{i}\}_{i=1}^{l}:\mathbf{H}=\Pi_{i=1}^{l}\mathbf{S}_{i} \Rightarrow\exists\{\mathbf{S}_{i}\}_{i=1}^{l+1}:\mathbf{H}=\Pi_{i=1}^{l+1}\mathbf{S}_{i}$. 

Consequently, to reach a fast method to compute $\mathbf{H}$, the error between $\mathbf{H}$ and $\Pi_{l}\mathbf{S}_{l}$ i.e. the error at each round can be minimized which can be stated as the following optimization problem
 \begin{subequations}\label{succgarph}
  \begin{align}
  \underset{\mathbf{S}_{1},\mathbf{S}_{2},\cdots,\mathbf{S}_{L}}{\text{min.}} &
         \sum_{l=1}^{L}{{\left\| \mathbf{H}-\Pi_{i=1}^{l}\mathbf{S}_{i} \right\|}_{F}^{2}}\label{2a}\\
\text{s. t.}   \;\;\;\;    	&  (\mathbf S_{i})_{n,n'}=0 \hspace{2mm} \nonumber\\&\text{if} \hspace{2mm} (v_n,v_{n'})\not \in \mathscr E ,  n,n' =1,....,N, i=1,...,L
\end{align}
\end{subequations}
To try to minimize the number of iterations, the weighted-sum method can be used. To do this, we accept larger amounts of error at early iterations to achieve smaller errors (which makes the method faster ) at the later iterations by assigning smaller weights to the error at earlier iterations, and larger ones to the error at last iterations. Therefore, one may think of assigning the weights increasingly. 

Thus, we have: 
 \begin{subequations}\label{final}
  \begin{align}
  \underset{\mathbf{S}_{1},\mathbf{S}_{2},\cdots,\mathbf{S}_{L}}{\text{min.}} &
         \sum_{l=1}^{L}\alpha_{l}{{\left\| \mathbf{H}-\Pi_{i=1}^{l}\mathbf{S}_{i} \right\|}_{F}^{2}}\\
\text{s. t.}   \;\;\;\;    	&  (\mathbf S_{i})_{n,n'}=0 \hspace{2mm} \nonumber\\&\text{if} \hspace{2mm} (v_n,v_{n'})\not \in \mathscr E ,  n,n' =1,....,N, i=1,...,L
\end{align}
\end{subequations}
where $\boldsymbol{\alpha}=[\alpha_1,\alpha_2,\cdots,\alpha_{L}]^{\top}$ is the weight vector whose entries are non-negative and $\sum_{l=1}^{L}\alpha_{l}=1$ w.l.o.g.

The optimization problem \eqref{final} is non-convex with respect to all the $\mathbf{S}_{1},\cdots, \mathbf{S}_{L}$, but it can be solved by the block coordinate descent (BCD) algorithm. In this approach, all variables are fixed except one of them, and the optimization problem is solved based on the variable (here the optimization problem with respect to each of variables is convex). This procedure repeats until all $\mathbf{S}_{1},\mathbf{S}_{2},\cdots,\mathbf{S}_{L}$ are considered as the variable of the optimization problem. The BCD algorithm repeats until a specified maximum number of iteration is attained.

 Consider the objective as a function of $\mathbf{S}_{j}$:

\begin{align}\label{cost}
J(\mathbf{S}_{j})=\sum_{l=j}^{L}\alpha_{l}{{\left\| \mathbf{H}-\Pi_{i=1}^{l}\mathbf{S}_{i} \right\|}_{F}^{2}}
\end{align}
By considering the topology constraint, we have:
\begin{align}\label{top}
J^{*}\overset{\Delta}=  \underset{\mathbf{S}_{j}\in\text{TOPF}}{\text{inf}} &
J(\mathbf{S}_{j})= \underset{\mathbf{S}_{j}\in\text{TOPF}}{\text{inf}}\nonumber\\&\sum_{l=j}^{L}\alpha_{l}{{\left\| \mathbf{H}-\mathbf{S}_{l}\mathbf{S}_{l-1}\cdots\mathbf{S}_{j+1}\mathbf{S}_{j}\mathbf{S}_{j-1}\cdots\mathbf{S}_{1} \right\|}_{F}^{2}}
\end{align}
where $\mathbf{S}_{j}\in\text{TOPF}$ means that $ (\mathbf S_{j})_{n,n'}=0 \hspace{2mm} \text{if} \hspace{2mm} (v_n,v_{n'})\not \in \mathscr E ,  n,n' =1,....,N$.  Furthermore, to make expressions shorter, we use $\mathbf{S}_{l:j}=\mathbf{S}_{l}\mathbf{S}_{l-1}\cdots\mathbf{S}_{j}$. By using the basis vectors, $\mathbf{S}_{j}$ can be expressed as follows:
\begin{align}
\mathbf{S}_j=\sum_{n,n'=1}^{N} S_{n,n'}^{(j)}\mathbf{e}_{n}\mathbf{e}_{n'}^{\top}
\end{align}
where $S_{n,n'}^{(j)}$ is the entry that lies in the $n$-th row and the $n'$-th column of $\mathbf{S}_j$. Then, we have:
\begin{align}
\mathbf{S}_{j}\in\text{TOPF}\iff\mathbf{S}_j=\sum_{(n,n')\in\mathscr E} S_{n,n'}^{(j)}\mathbf{e}_{n}\mathbf{e}_{n'}^{\top}
\end{align}
Consequently, \eqref{top} can be rewritten as:
\begin{align}\label{mat}
J^{*}=  \underset {s.t.\mathbf{S}_{j}=\sum_{(n,n')\in\mathscr E}  S_{n,n'}^{(j)}\mathbf{e}_{n}\mathbf{e}_{n'}^{\top}}{\text{inf}} &
J(\mathbf{S}_{j})= \nonumber\\\underset{\{{S}^{(j)}_{n,n'}\}_{n,n'\in\mathscr E}}{\text{inf}}&J(\sum_{n,n'\in\mathscr E}^{N} S_{n,n'}^{(j)}\mathbf{e}_{n}\mathbf{e}_{n'}^{\top})
\end{align}
Now, the variables of \eqref{mat} are the entries of $\mathbf{S}_{j}$ which can be expressed by vectorized form of $\mathbf{S}_{j}$. Therefore, by applying the vectorization operator and $\mathrm{vec}(\mathbf{e}_{n}\mathbf{e}_{n'}^{\top})=\mathbf{e}_{n'}\otimes\mathbf{e}_{n}$, the term inside of $J$ ($\sum_{n,n'\in\mathscr E}^{N} S_{n,n'}^{(j)}\mathbf{e}_{n}\mathbf{e}_{n'}^{\top}$) is expressed as follows: 
\begin{align}\label{vec}
\mathrm{vec}(\sum_{(n,n')\in\mathscr E} S_{n,n'}^{(j)}\mathbf{e}_{n}\mathbf{e}_{n'}^{\top})=\sum_{(n,n')\in\mathscr E} S_{n,n'}^{(j)}(\mathbf{e}_{n'}\otimes\mathbf{e}_{n})
\end{align}
Then
\begin{align}\label{vec1}
\sum_{(n,n')\in\mathscr E} S_{n,n'}^{(j)}(\mathbf{e}_{n'}\otimes\mathbf{e}_{n})=\underbrace{[\mathbf{e}_{n'_{1}}\otimes\mathbf{e}_{n_{1}},\cdots, \mathbf{e}_{n'_{E}}\otimes\mathbf{e}_{n_{E}}]}_{\mathbf{A}}{\mathbf{s}^{(j)}}
\end{align}
where $\mathscr {E}=\{(n_1,n'_1),\cdots, (n_E,n'_E)\}$ and $\mathbf{s}^{(j)}=\left[\begin{array}{ccc}
   S_{n_1,n'_1}^{(j)} &
    \cdots &
   S_{n_E,n'_E}^{(j)} 
\end{array}\right]^{\top}$. Thus, \eqref{mat} can be stated as follows:
\begin{align}
 \underset{\{S^{(j)}_{n,n'}\}_{n,n'\in\mathscr E}}{\text{inf}}&J(\sum_{n,n'\in\mathscr E}^{N} S_{n,n'}^{(j)}\mathbf{e}_{n}\mathbf{e}_{n'}^{\top})=\underset{\mathbf{s}^{(j)}\in\mathbb{R}^{E}}{\text{inf}}&J(\mathrm{vec}^{-1}(\mathbf{A}\mathbf{s}^{(j)}))
\end{align}
Consequently, \eqref{succgarph} can be stated as
\begin{align}\label{fin}
\underset{\mathbf{s}^{(j)}}{\text{inf}} &
         \sum_{l=j}^{L}\alpha_{l}{{\left\| \mathbf{H}-\mathbf{S}_{l:j+1}\mathrm{vec}^{-1}(\mathbf{A}\mathbf{s}^{(j)})) \mathbf{S}_{j-1:1}\right\|}_{F}^{2}}
\end{align}

Finally, we can write \eqref{fin} in vectorized form as follows:
\begin{align}\label{fin1}
\underset{\mathbf{s}^{(j)}}{\text{inf}} &
         \sum_{l=j}^{L}\alpha_{l}{{\left\| \mathrm{vec}(\mathbf{H})-(\mathbf{S}_{j-1:1}^{\top}\otimes\mathbf{S}_{l:j+1})\mathbf{A}\mathbf{s}^{(j)}\right\|}_{2}^{2}}
\end{align}

The following algorithm indicates the approach for solving \eqref{fin1}.  
\begin{algorithm}[h!]
\caption{Proposed solver}
\label{my_algorithm}
\begin{algorithmic}[1]
\Require  $ I_\text{MAX}$.
\State{set $\mathbf{S}_{2}, \mathbf{S}_{3},\cdots, \mathbf{S}_{L}$ as the identity matrix}

	\For{$i = 1$ to $I_\text{MAX}$}
\For{$j= 1$ to $L$}
\State{fix $\mathbf{S}_{m},  m=1,\cdots, L, m\neq{j}$ and obtain $\mathbf{s}^{(j)}$ by solving \eqref{fin1}}	
	\EndFor
\EndFor
\State \Return	 $\mathbf{S}_{1}, \mathbf{S}_{2},\cdots, \mathbf{S}_{L}$				
\end{algorithmic}
\end{algorithm}

\noindent where $I_\text{MAX}$ denotes the maximum number of iterations of the BCD algorithm. 

\section{Conclusion}
This paper proposes a decentralized algorithm to design the graph shift operators for obtaining linear transformation functions in a nearly minimal number of iterations. The approach relies on a sequence of graph shift operators, and tries to minimize the number of iterations needed to design linear transformation functions.
\bibliographystyle{ieeetr}
\bibliography{refs}
\end{document}